# Programmable CMOS DAC Operating in Cryogenic Environments for Controlling Superconducting Qubits

Northrop Grumman Microelectronics Design and Applications*

Northrop Grumman Systems Corporation

880 Elkridge Landing Rd, Linthicum Heights, MD 21090

This paper presents the design and test results of a CMOS current-based Digital to Analog Converter (DAC) that operates at cryogenic temperatures and that can be used to precisely control the amount of flux coupled to qubits or that can be used in the readout of superconducting circuits. The current pulse output can be controlled in terms of its amplitude, rise and fall slopes, via digital controls, and pulse width, via external triggers, while driving a superconducting circuit. The design has been implemented in a planar 90nm CMOS process and test results closely match circuit predictions. The solution's wide degree of digital tunability affords potential application of the same device to many types of quantum circuits, beyond those discussed here. Due to the wide-ranging flexibility of digital CMOS control, we envision that this DAC design will enable the next generation of high-fidelity cryo-CMOS control architectures for superconducting qubits.

## I. INTRODUCTION.

Superconducting qubits [1] [2], [3] are one of the current leading industry standards for building scalable quantum processing units [4], [5] owing to advantages in speed [6], [7], and manufacturing maturity [8], [9] with the promise to solve the wiring bottleneck by incorporating either superconducting digital logic [10], [11], [12] or cryo-CMOS control electronics [13], [14], [15], [16] . Superconducting digital logic [17] presents a scalable alternative to microwave-controlled dispersive readout [18], [19], the main readout method that's currently used for superconducting qubits. Here, we present a new way to perform flux-controlled superconducting readout with CMOS control.

The goal of this work is to generate analog waveforms which control the amount of flux in a superconducting loop via mutual coupling. The DAC was designed for high-fidelity control of a superconducting flux qubit and quantum flux parametron (QFP) $\alpha$ loops (Figure 1) and can be rapidly turned on/off. Additionally, this design has broader applicability to controlling different types of quantum circuits through its digitally-tunable waveform slew rate and amplitude.

The cryo-CMOS control architecture that we envision involves either directly bump-bonding the control chip to the superconducting quantum chip, or to an interposer chip. The latter may be necessary in order to shield the quantum chip from thermal radiation, insulate the quantum chip through electrically-conductive thermal standoffs, and/or to house control noise filters. One advantage of flux qubit readout is that the persistent current in the QFPs is large enough that it can be readily converted back into a robust digital signal using a CMOS threshold detector. The DAC presented here has been demonstrated to perform high-fidelity qubit readout and preparation in simulation, and is believed to also be highly-suitable for controlling two-qubit gates. In order to perform single-qubit gates, one would need to design an additional, similar CMOS DAC that's faster and lower amplitude. Once combined together onto a single CMOS chip, these three components would form a noisy-intermediate-scale-quantum control device that has the potential for scalability.

Two iterations of the CMOS current DAC were designed and fabricated: iteration 1 was designed in the SkyWater® S90ROIC process, while iteration 2 was designed in GlobalFoundries® 9HP+; both are 90nm planar processes. Iteration 2 included more configuration bits for both pulse amplitude and ramp up/down rates, giving it a wider range of amplitude control as well as finer granularity tuning of the ramp rate. Design iteration 1 was simulated and fully tested at both room temperature and cryogenic temperatures ranging from 4 K to 600 mK. Design iteration 2 was fabricated at a later date and had yet to be tested as of the writing of this paper (though simulation data has been included here). Although design iteration 1 will be the focus of this paper, design iteration 2 is expected to provide two orders-of-magnitude improved gate fidelities as shown in Figure 2.

## II. DESIGN APPLICATIONS.

A digitally programmable CMOS current DAC can be used to inject a well-controlled amount of flux [20] into a loop of a flux-controlled superconducting qubit, such as a Transmon [6], flux qubit [21] [22], Fluxonium [23], or quantum gate couplers [24]. One possible application is shown in Figure 1.

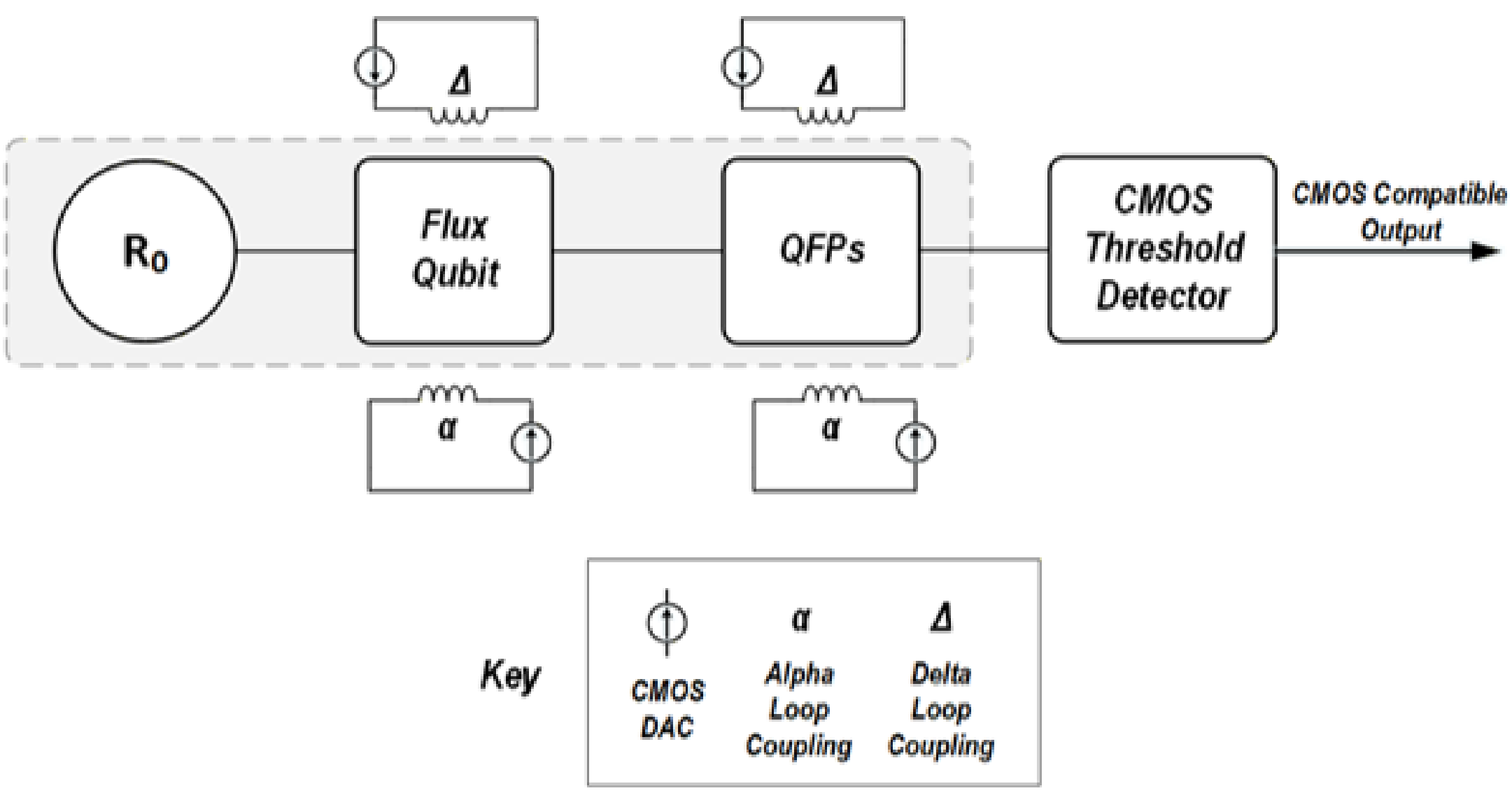

Figure 1 – The application of this fast IDAC. R0 is a superconducting qubit. DC controls are applied to bias the QFPs near its operating point, and a fast flux pulse provided by this DAC controls its operation.

The preparation and readout fidelity of a qubit is limited by intrinsic decoherence, diabatic transitions, GHz-range spectral content in the control waveform, and GHz-range noise on the control lines. These factors translate to the following respective requirements on the control waveform: minimal regions of constant flux during ramping, digital-tunability of slew rate, smoothness during ramping, and post-DAC filtration. The CMOS DAC produces an analog waveform without step-like features, which is a limitation of single flux quanta (SFQ) DACs [11]. The second iteration CMOS DAC eliminates delays at the beginning and end of the waveform, allowing for a slower slew rate during regions where diabatic transitions can occur. For fixed preparation and readout times, this results in a lower simulated diabatic transition probability than the first iteration DAC, as shown in Figure 2.

Quantum simulations of the CMOS DAC-controlled qubit preparation and readout were performed using Northrop Grumman's proprietary superconducting circuit quantum simulator, Circuitizer. Here, we define leakage error as the probability to be outside of the target state after preparation or the initial state after readout. For the DAC simulations, we turned off the intrinsic decoherence of the superconducting qubit, since our goal was to assess the control errors. The waveform-induced leakage error in the absence of control noise is shown in Figure 2, and the effect of nominal control noise combined with a post-DAC filter is shown in Figure 3. The actual control noise experienced by the superconducting qubit from the CMOS DAC depends on the temperature associated with the self-heating of the CMOS circuit during operation, along with thermal routing on chip, and the fridge cooling power.

We modeled the control noise temperature as nominal Johnson-Nyquist noise with resistance R = 50 Ω and temperature T = 1 K. The quantum state leakage errors scale with the noise power as ∝ T/R. Thus, once the CMOS noise temperature is known more precisely [25], our results can be rescaled to the appropriate noise temperature accordingly. Figure 3 shows that even T = 1 K control noise will result in untenably large excitation probabilities of a superconducting qubit. However, a first-order Butterworth low-pass filter can eliminate the effects of control noise, in simulation, once the cutoff frequency is below the lowest-energy relevant quantum transition frequency. The relevant transition frequencies are those between the occupied quantum state and other appreciably-coupled modes during the qubit preparation or readout process.

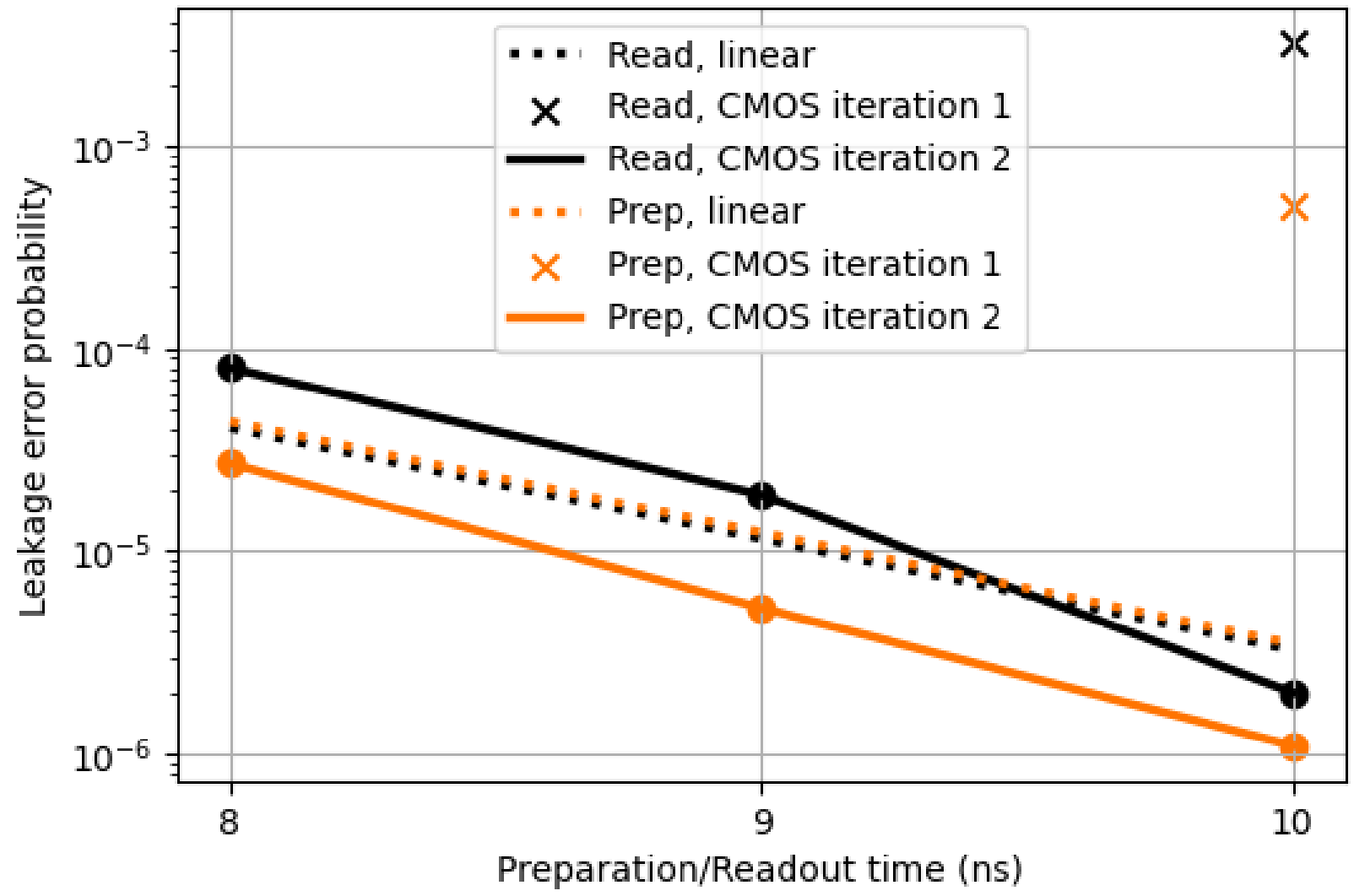

Figure 2 – Noise-free Circuitizer simulation of the readout and preparation fidelity of a superconducting qubit when controlled by the CMOS current DAC. The second iteration device reduced delays at the beginning and end of the ramps, enabling higher fidelity for the same total gate time.

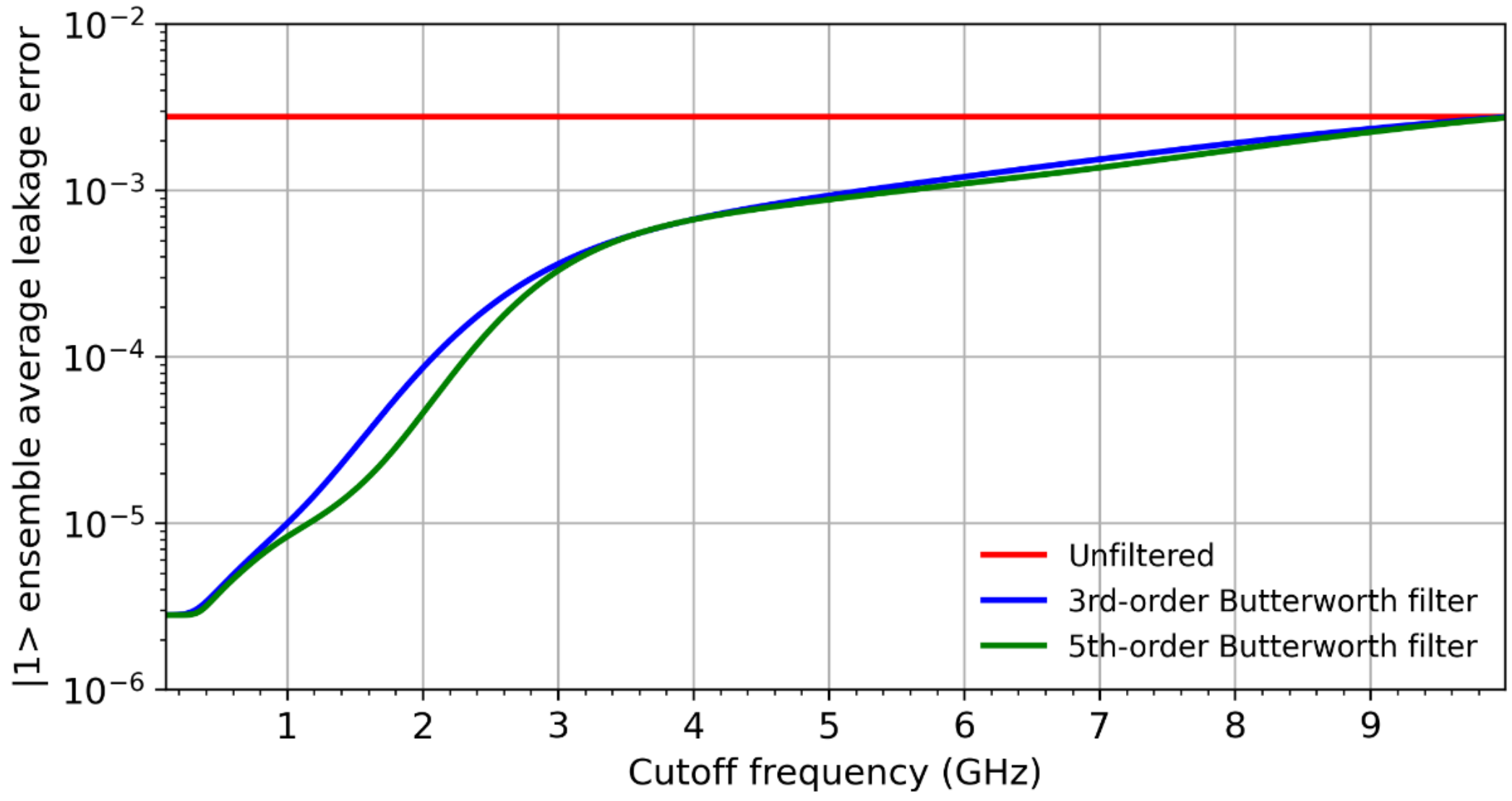

Figure 3 – CMOS DAC iteration 1 readout fidelity when there is nominal R = 50 Ω, T = 1 K thermal noise on the CMOS DAC control line (red), and when there is a low-pass filter attached to the output (green, blue). The low-pass cutoff frequency is swept on the x-axis, showing that the fidelity returns to its noise-free value once the noise PSD at frequencies above the minimum, relevant quantum transition frequency is filtered out.

Filtering can be incorporated primarily off-chip to generate smoother analog waveforms. On-chip filtering is limited due to area/power restrictions and off-chip filtering can be better targeted in terms of lower frequency operation. Additionally, the CMOS current DAC could be used to drive a superconducting circuit with inherent control noise filtration, before ultimately driving a superconducting circuit, such as a flux qubit.

## III. DIGITALLY-PROGRAMMABLE CMOS CURRENT DAC.

### A. Functional Description.

The amount of flux injected into a superconducting loop is primarily controlled via a digitally controlled current pulse which is depicted in Figure 4, below; the current pulse is initiated and subsequently ends following a nominal delay, τ, after the start and stop triggers are received, respectively. The current pulse features linear rise and fall slopes that are varied using digital controls as well as a digitally programmable amplitude.

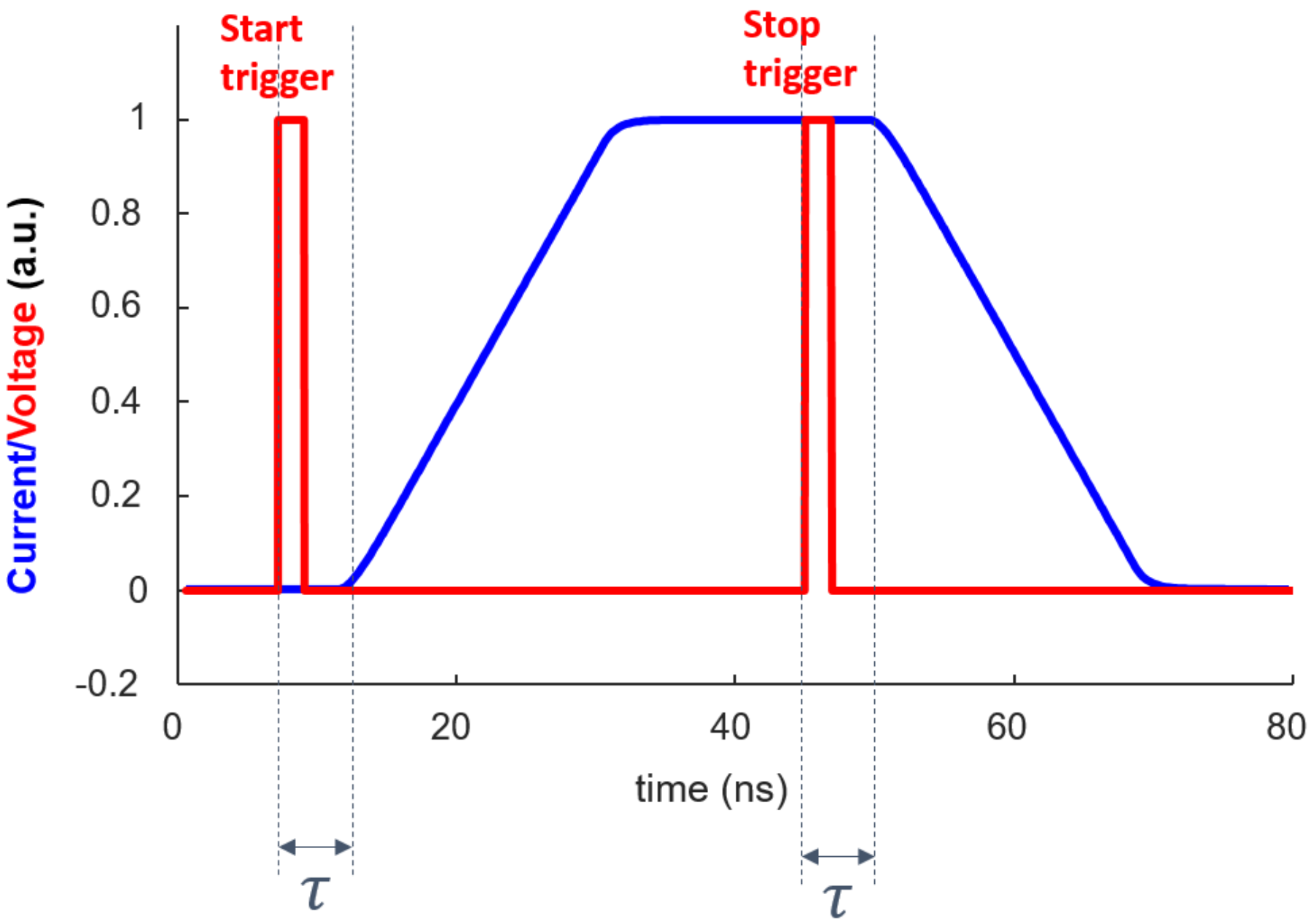

Figure 4 – Functional Depiction of the Current Pulse, showing ideal operation of the device. An initial "start" signal is sent to the device, which begins its output current ramp up after some time τ. The output current ramps down some time τ after receiving a "stop" signal.

The programmable current pulse is generated via a CMOS current DAC, which is shown in Figure 5. The design utilizes a digital block for providing all the necessary control bits and a bias control module for appropriate biasing of all internal circuits. In addition, a signal conditioning module takes in external stimuli, such as a reset signal, a low power mode enable signal, as well as start and stop triggers, – these are used for reset, distributing the low-power mode enable signal to different blocks, and controlling the start and stop of the output current waveform. The rise and fall times of the output current waveform are digitally tunable via the slope-control module. The rise and fall slopes of the output waveform are varied digitally by controlling the charge and discharge rates of a thermometer-coded capacitor bank via a binary weighted current DAC. The capacitor voltage is subsequently converted to current and the output amplitude is tuned using a 5-bit, binary-weighted current DAC. This design was fabricated using a 90nm planar CMOS process and has been tested in a laboratory.

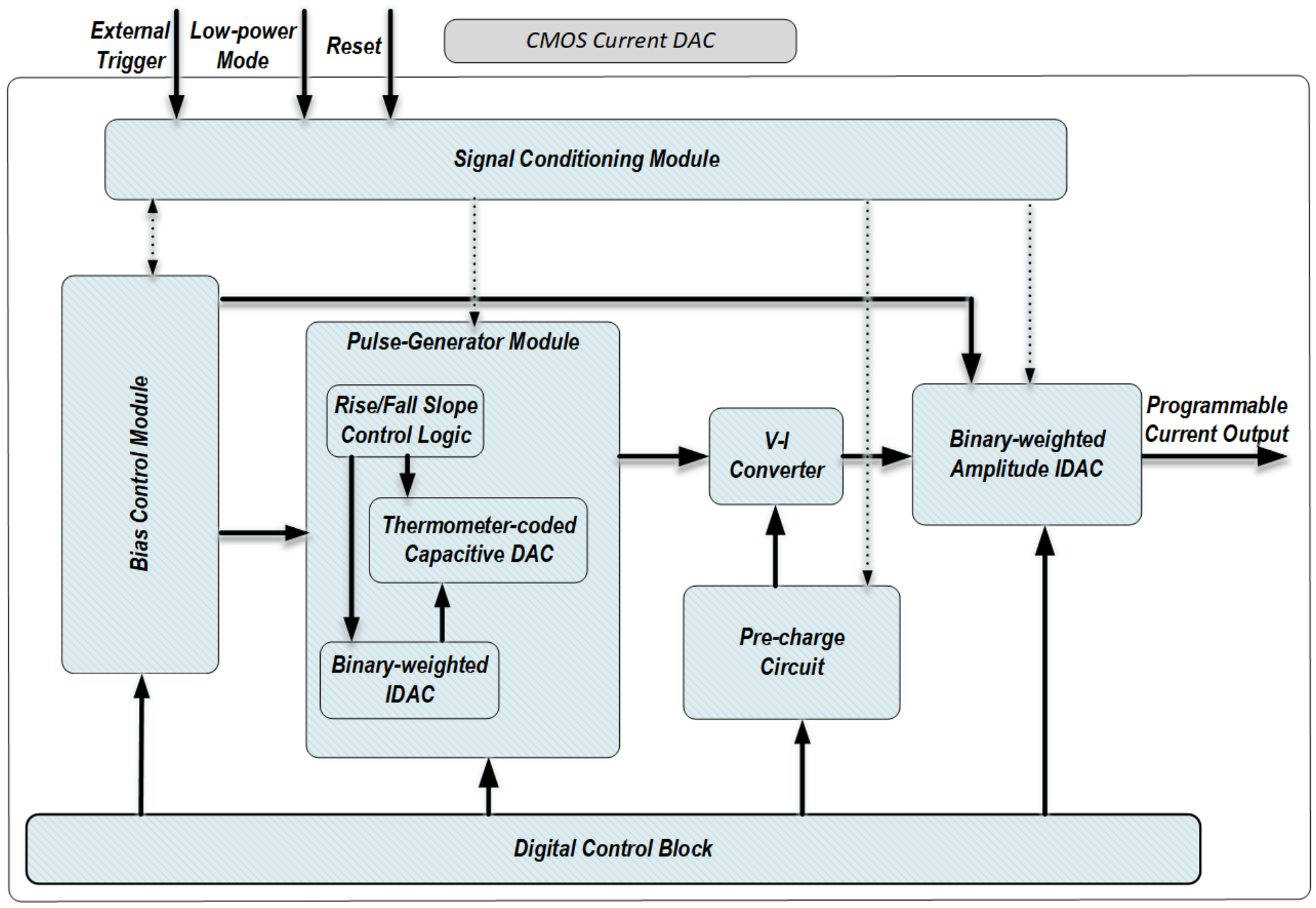

Figure 5 – Overall block diagram of the CMOS current DAC. Here the external trigger signal refers more generically to both the start and stop triggers that are used for initiating the start and stop of the output current waveform.

The CMOS current DAC design minimizes the delays associated with the start of the output waveform by utilizing a novel pre-charge circuit. The pre-charge level is digitally tunable which additionally improves the dependence of the waveform start on circuit and parameter variations. The pre-charge circuit uses tri-state inverters connected to a bank of binary-programmable capacitors in order to adjust the input pre-charge voltage to the V-I converter, minimizing simultaneously the leakage current and the start delay of the output current pulse. A simplified block diagram of the pre-charge circuit is shown in Figure 6, below; the diagram includes the truth table for the tri-state inverter. In this design, four-bit selects are used to control the tri-state inverter of each sub-circuit, all of which contain binary-weighted capacitor banks C(m) – the number of control bits can be easily expanded, as needed, in the future. The charge select and evaluate signals are non-overlapping clock pulses generated from an external trigger (in this case the start trigger is used) and distributed to each sub-circuit. Finally, the output is gated by the pre-charge enable control bit which can enable or disable the pre-charge circuit output altogether.

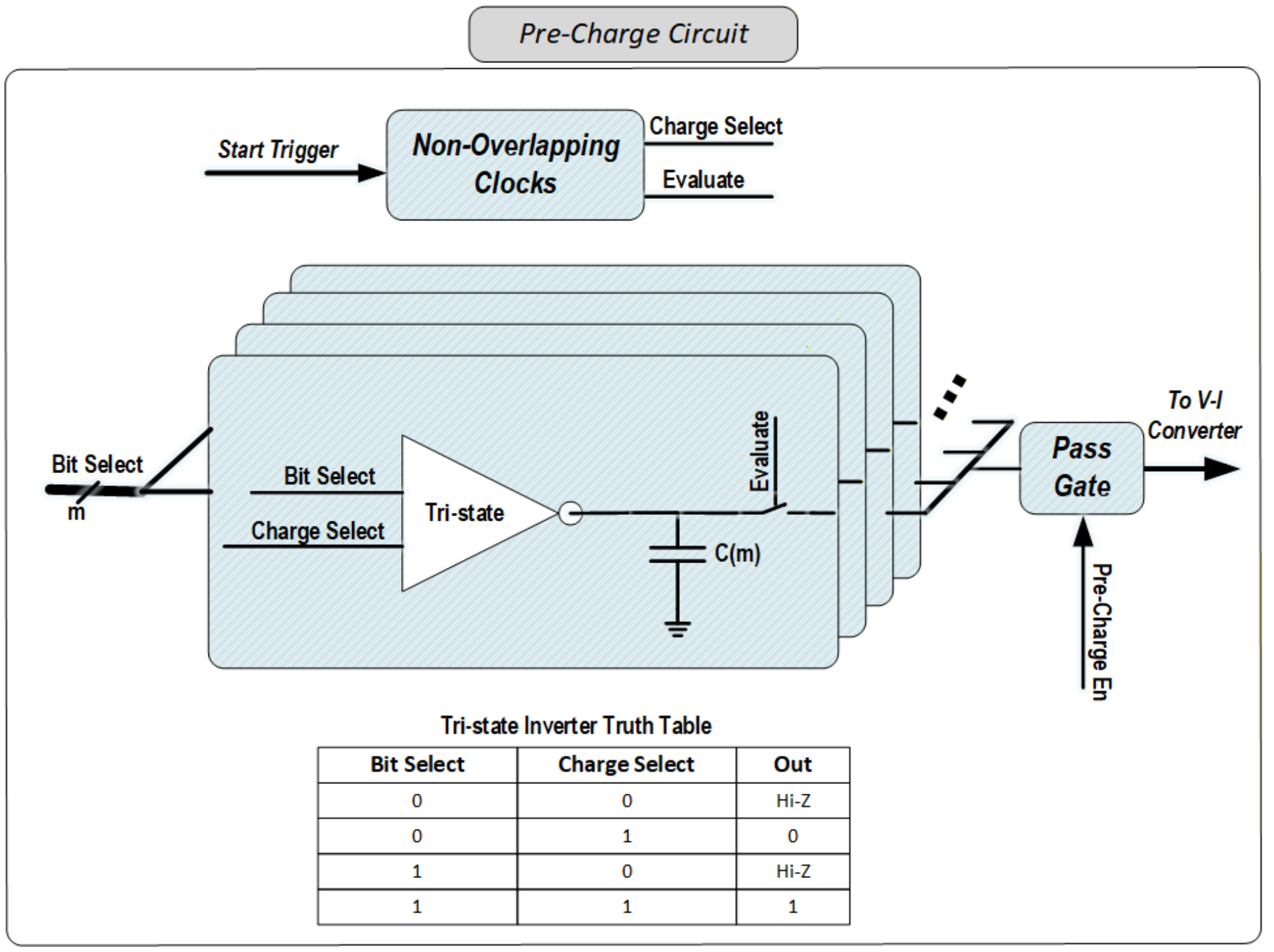

| Bit Select | Charge Select | Out |
|---|---|---|
| 0 | 0 | Hi-Z |
| 0 | 1 | 0 |
| 1 | 0 | Hi-Z |
| 1 | 1 | 1 |

Figure 6 – Block diagram of Pre-Charge Circuit, demonstrating how capacitor banks are selected by signals from the digital control logic.

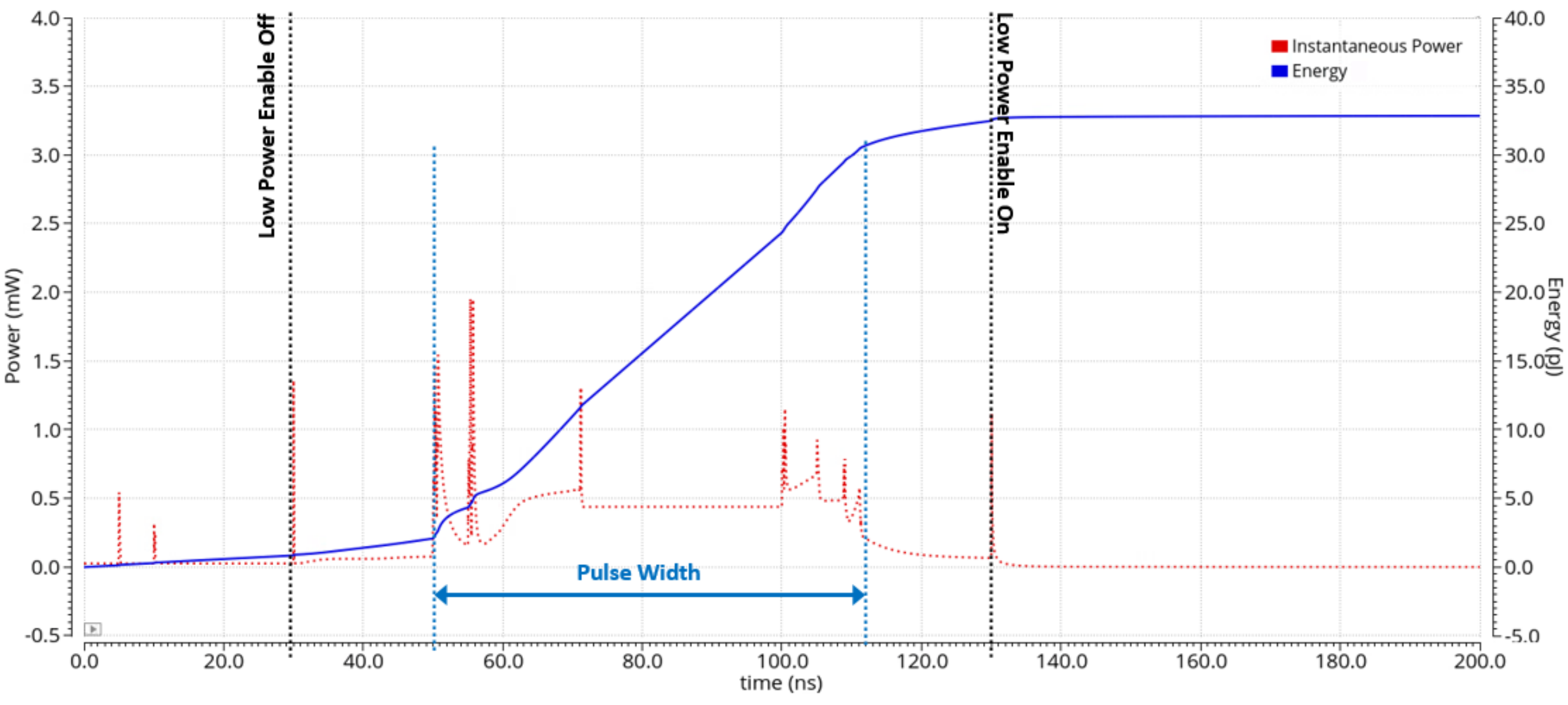

Figure 7 – Representative Instantaneous Power and Energy per Preparation and Readout Operation.

Figure 7 shows the total instantaneous power dissipated by the CMOS current DAC during a preparation and readout operation with a pulse width of 60ns, using a pulse amplitude of 190 uA, and rise and fall times of approximately 8 ns. Total energy for this operation is approximately 32.5 pJ, primarily dominated by the power dissipated during the ramp-up, hold, and ramp-down of the output pulse, and quickly flattens as the low-power mode is turned on.

### B. Circuit Simulations with Cryogenic Models.

The DAC design has been simulated using Spectre®. Simulation results for the output amplitude and rise and fall times of iteration 1 of the device are shown in Figure 8a and Figure 8b, respectively; the simulations sweep the control digital words (separately) for the amplitude and rise/fall time - here the 10-90% rise and fall time standard definitions are used. The circuit avoids ringing/overshoot for a reasonable load inductance range ($L_{Load}$< 100 nH), has amplitude flatness of <1%, and also has a delay, $\tau$, of less than 20 ns. Figure 8c and 8d show the same sweeps for iteration 2 of the device, which improves the tuning ranges of the amplitude as well as the rise and fall times, which ultimately results in better flux control. The figure below highlights the main differences in the simulated responses between the two design iterations; the rest of the paper will focus solely on design iteration 1.

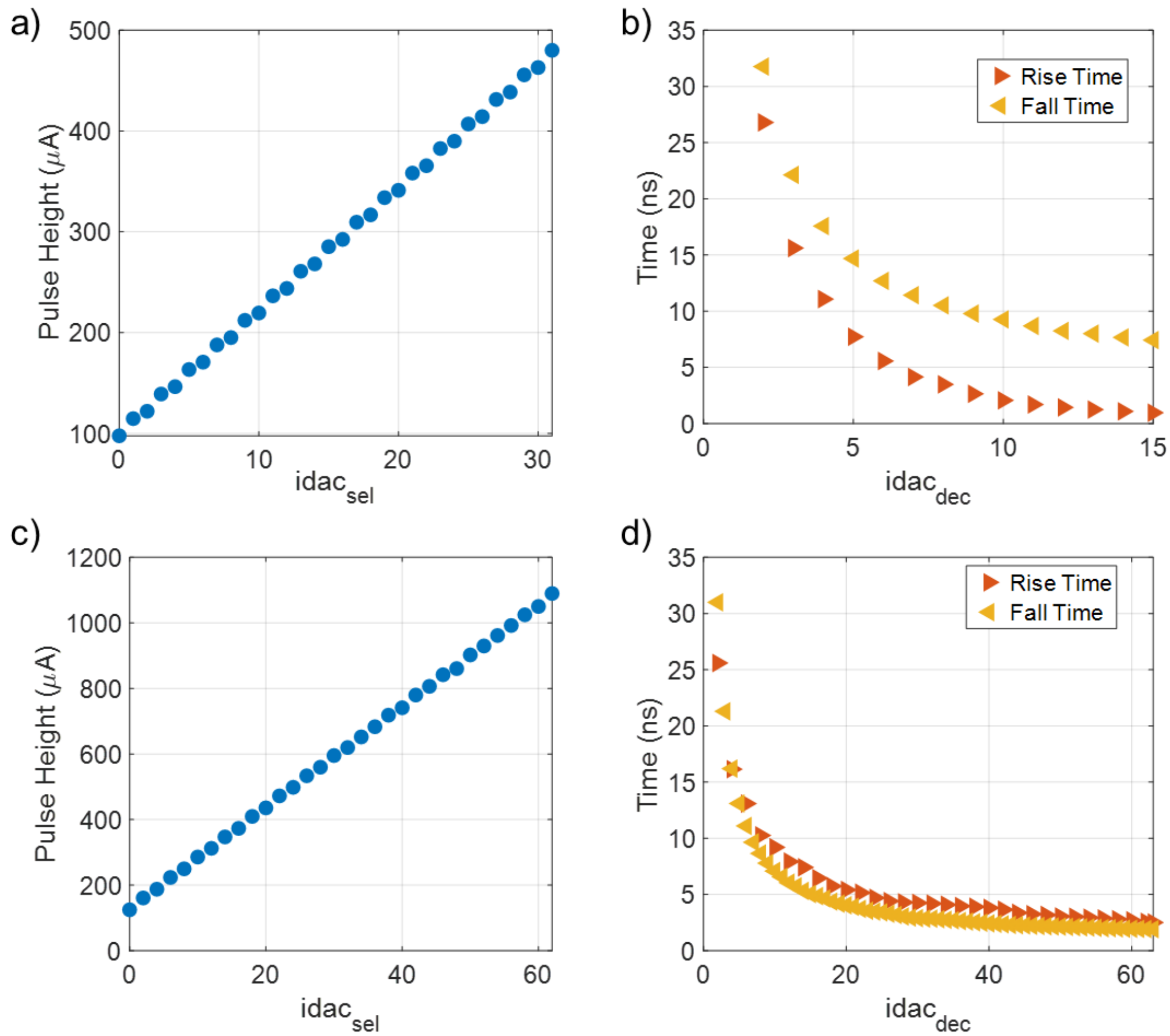

Figure 8 – a) Simulated CMOS IDAC iteration 1 Current Amplitude Sweep at Cryogenic Temperature. b) Simulated CMOS IDAC iteration 1 rise and fall time sweep at Cryogenic Temperature. c) Simulated CMOS IDAC iteration 2 Current Amplitude Sweep at Cryogenic Temperature. d) Simulated CMOS IDAC iteration 2 rise and fall time sweep at Cryogenic Temperature.

The CMOS current DAC has also been simulated over process corners using a nominal output current of around 190 uA; the results are recorded in Figure 9. The pre-charge circuit was not enabled during these simulations to expedite the results. Two waveforms are reported per corner: one with digital corrections applied and one without - traces closer in amplitude to the nominal include digital adjustments. As seen from the plot, there are large variations observed in the output amplitude; the current amplitude varies from approximately 50% to 200% of the target value. The results further underscore the need for digital tunability (more control bits were included in a second iteration of the design) albeit the results tend to be pessimistic, especially in the slow corner case. Since device mismatch degrades at cryogenic temperatures [26], the amplitude variation was simulated using the built-in Spectre® Monte Carlo tool with 250 runs; the resulting Gaussian-like distribution can be seen in Figure 10.

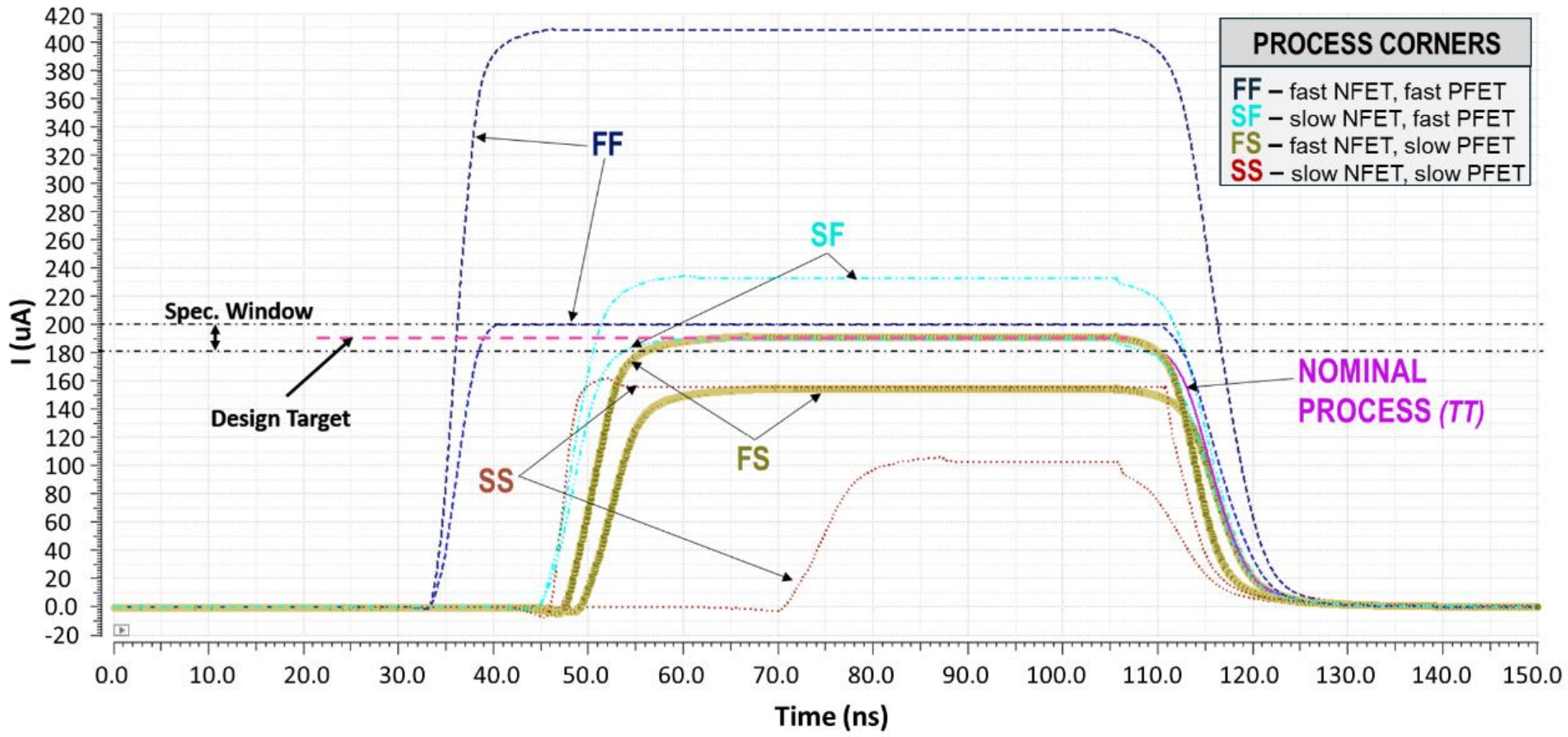

Figure 9 – Simulated CMOS current DAC over corners plus nominal. Traces closer in amplitude to the nominal include digital adjustments; the nominal trace is not entirely visible due to overlap with the fs corner.

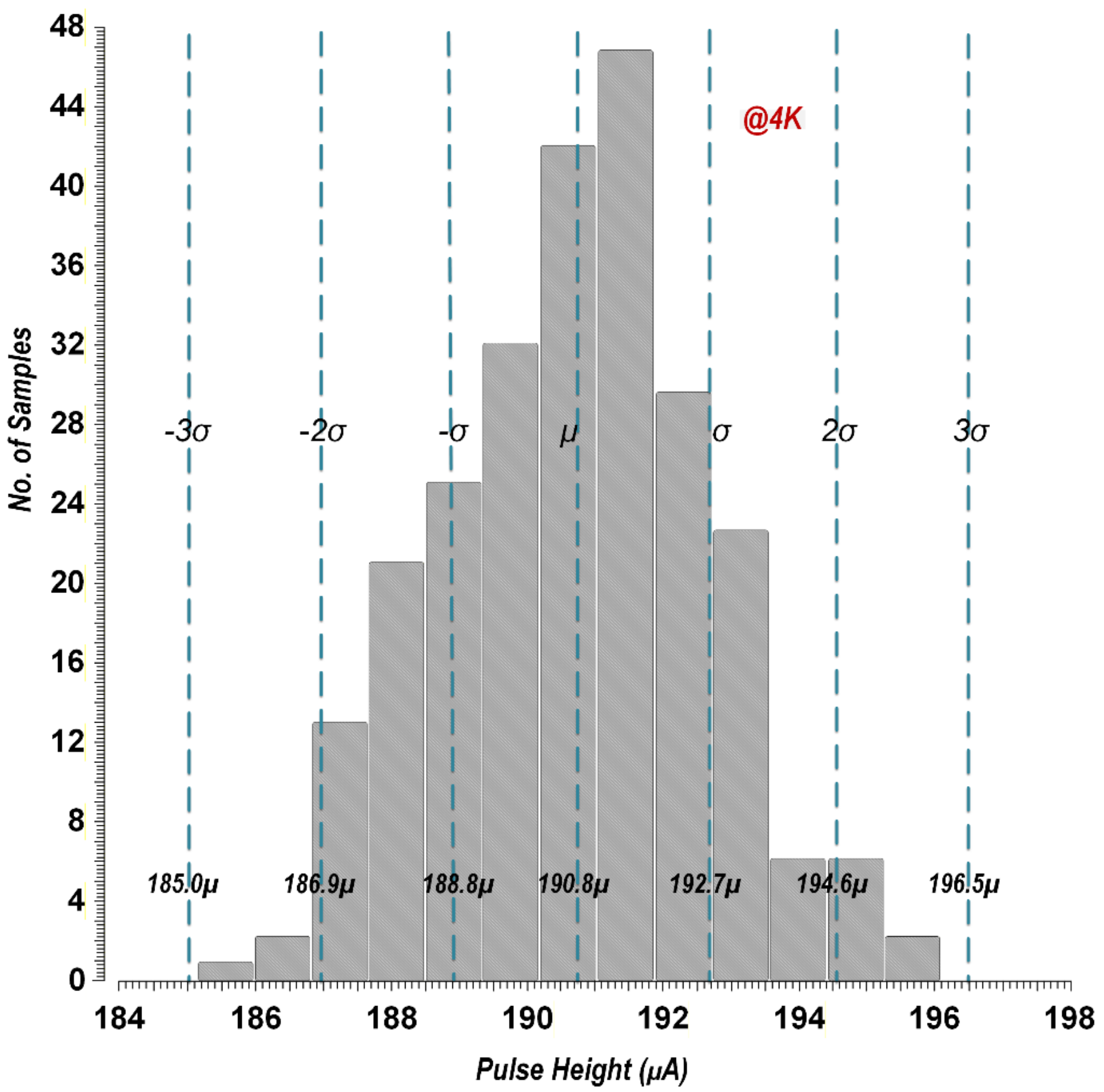

Figure 10 – Monte Carlo simulation of the CMOS DAC output for a targeted current value of around 190uA. The Gaussian-like distribution has a standard deviation of approximately 2µA.

Overall, Spectre® simulations with cryogenic models show that the standard CMOS components are working within the expected range. The device may perform somewhat differently when measured at cryogenic temperatures (e.g., lower maximum amplitude, brief waveform delays, etc.), but is expected to be similar. Furthermore, these differences are unlikely to significantly affect the ultimate device performance due to the device's wide degree of digital waveform tunability.

### C. Overview of the Digital Design Wrapper and Chip Integration.

The CMOS current DAC is digitally controlled by two separate layers of digital control. All devices in the digital control layers use a supply and ground separate from that of the current DACs, for purposes of isolating the DACs from the switching noise of the digital devices. The first layer consists of a resettable finite state machine (FSM) that determines which of the on-die DACs will be exercised, or if the chip will be in a top-level configuration state. The reset state of the FSM places it in the top-level configuration state, which uses a SPI-style (serial peripheral interface) scheme to take serial data from room temperature into the least significant bit (LSB) of a shift register, which is clocked by another signal provided from room temperature. The most significant bit (MSB) of the shift register sends data back up to room temperature to verify operation and control of the chip. When a signal from a trigger signal line is received, each bit in the shift register delivers data in parallel to a resettable buffer register array of equal size, the outputs of which are connected to the ports that configure the IO cells. This scheme effectively shields the configuration ports of the IO cells from bits shifting across the shift register, preventing loss of control of the chip during configuration. The remaining states of the FSM control which DAC experiment is selected, and re-routes room temperature signals and room temperature DC current bias lines to the selected experiment. The signal lines that are re-routed using this scheme are the serial data entry line, the clock line for sequential elements, the reset line for sequential elements, and a pulse start signal. The trigger line from the reset state of the FSM is also re-routed and repurposed as a pulse stop signal for the experiment selection states. These re-routed signal lines are received by the second layer of digital control.

The second layer of digital control consists of a set of digital wrappers individual to each DAC that control the delivery of trim configuration data to the selected DAC, and directly control the operation of the output pulse. The digital interface of the wrappers is also SPI-based, utilizing a data entry shift register that takes serial data from room temperature at its LSB, and reports serial data back to room temperature from its MSB. The MSB reporting from the shift register is again used to verify proper operation of the shift registers in the design. Due to line count limitations, the trigger line method used to deliver data from the shift register cannot be used, and a code-word scheme is used instead. To accomplish this, the shift register is broken into two major sets of addresses: [n-1:0], where n is the number of DAC configuration ports, and [N-1:n], where N is the total number of bits in the shift register. Each bit in the address space [n-1:0] corresponds to a unique configuration port on the current DAC that controls its tunability. Address space [N-1:n] delivers data from each bit to a code-word comparator. When the correct code-word occupies addresses [N-1:n], the data from addresses [n-1:0] is flushed into a resettable buffer register array of equal size. The outputs of the buffer array registers deliver data signals to the DAC configuration ports. Much like the top-level control of the IO cells, this code-word and buffer scheme was implemented in order to mask the DAC configuration ports from bits moving through the data entry shift register during experimental setup. Control of the output pulse is directly handled by a pulse start signal and a pulse stop signal. When the pulse start signal is held high for 10 ns, the DAC ramps up its output current, and when the pulse stop signal is held high for 10 ns, the DAC output current ramps down.

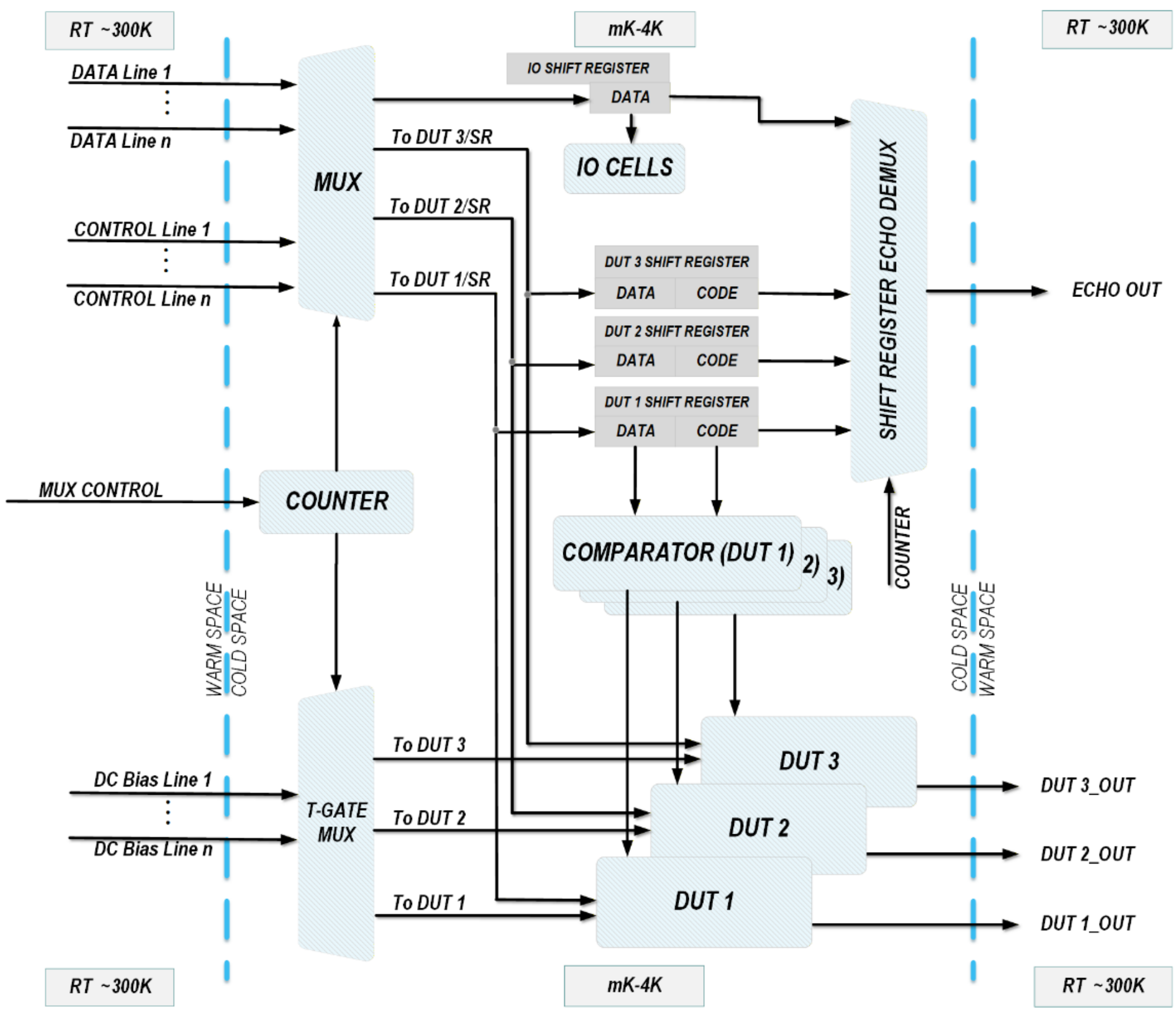

Figure 11 – Digital block level diagram, showing how signals are routed from room temperature to the various DUTs (devices under test) on die, and sent back to room temperature.

Three different experiments, also denoted as MUX states, were included at the top level; these experiments included identical CMOS current DAC devices with 50 Ω, return-current terminations. The output current pins connect to superconducting loops with different characteristics, for example different flux qubit α-loops, while the current-return 50 Ω transmission lines terminate on chip as annotated in Figure 12, below. The CMOS DAC Application Specific Integrated Circuit, or simply ASIC, measures 5mm x 5mm, and it includes the on-chip digital core, the digital I/O on the left and right sides, bias control pins on the south side, and connections to superconducting circuits on the north side.

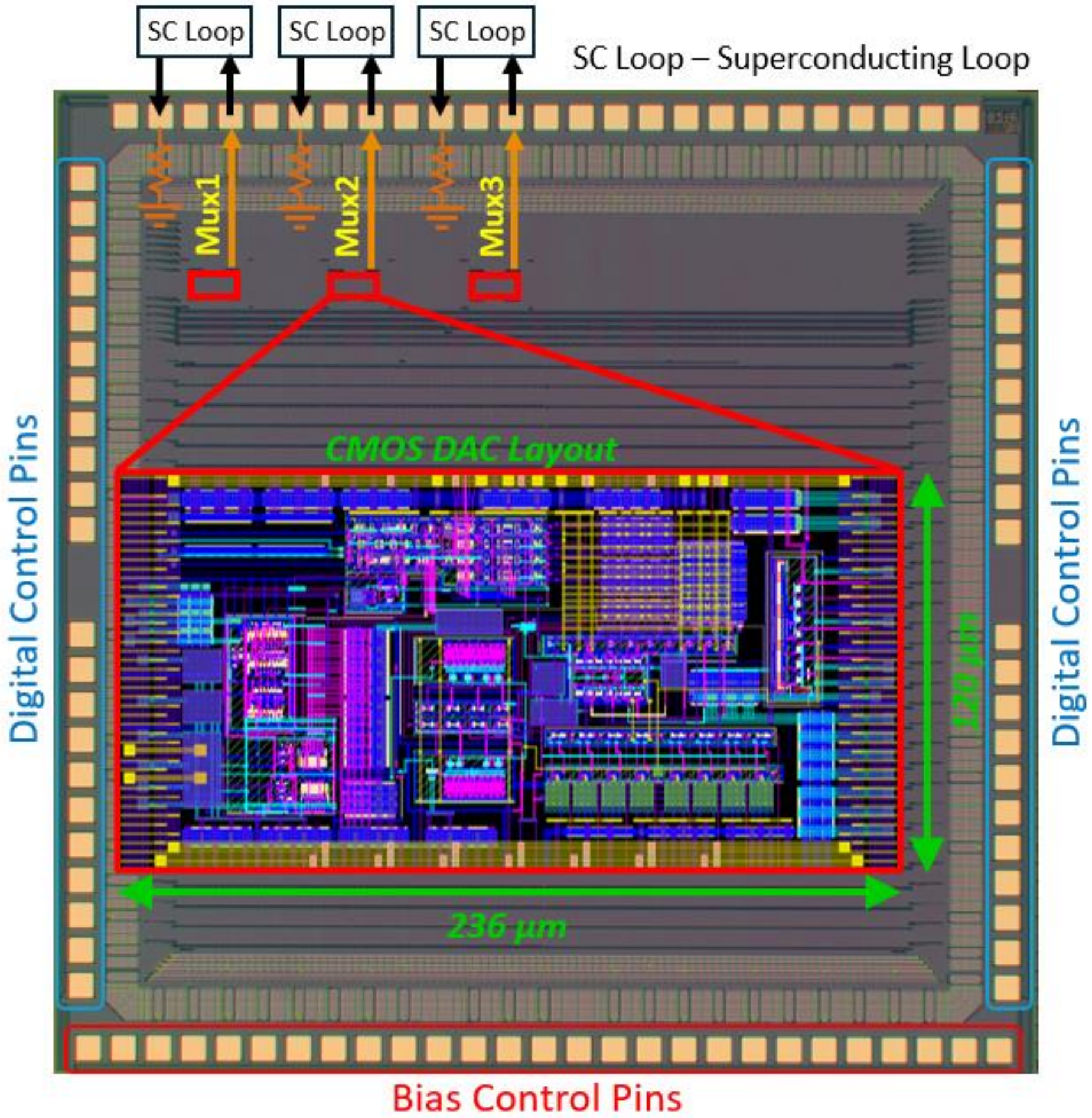

Figure 12 – Chip micrograph of the 5 mm x 5 mm ASIC with overlaid layout of the CMOS current DAC. Three experiments (denoted as Mux states) are included per chip and encircled above.

## IV. TEST RESULTS.

Test results at room temperature and at 600 mK are presented and discussed below. It is worthy to note that there is no particular significance to 600 mK, this was arbitrarily chosen to observe the operation of the ASIC at some cryogenic temperature; in testing unreported here the chip operated very similarly at 4 K. The chip is expected to be operational at temperatures between 600 mK and 300 K, although only room temperature and 600 mK results were recorded and are reported in this paper.

## A. Room Temperature Test Configuration and Measured Results.

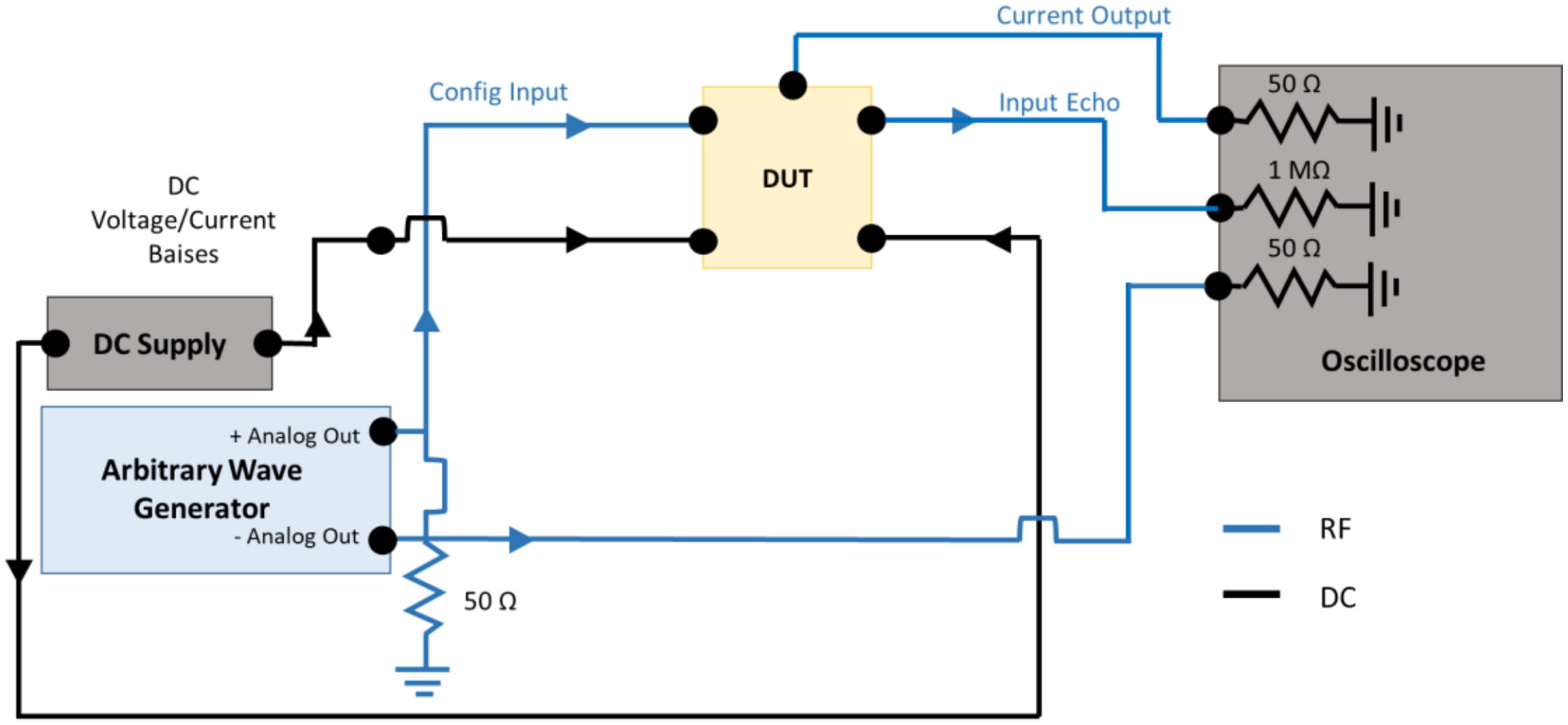

Figure 13 – High-level overview of the room temperature test setup.

The analog output functionality was first measured on a room temperature test bench, as shown in Figure 13, prior to testing it in a dilution refrigerator. The test bench uses a Tektronix® AWG 5208 arbitrary waveform generator (AWG) for sending the clock signal, data sequences, and trigger pulses to the ASIC device under test (DUT). The DC biases were held using a Keithley® 2450 low noise DC power supply. For the room temperature tests, the digital block was clocked at 1.25 MHz, and triggers were 10ns wide. All AWG signal amplitudes were set to match the IO supply voltage of 1.8 V and were terminated with 50 Ω external resistors. The AVDD and DVDD voltages were set to 1.2 V. Analog outputs were measured via a Tektronix® MDO34 multi-domain oscilloscope with internal 50 Ω termination. The cable lengths for the room temperature test setup were 12 inches between the equipment and the chip.

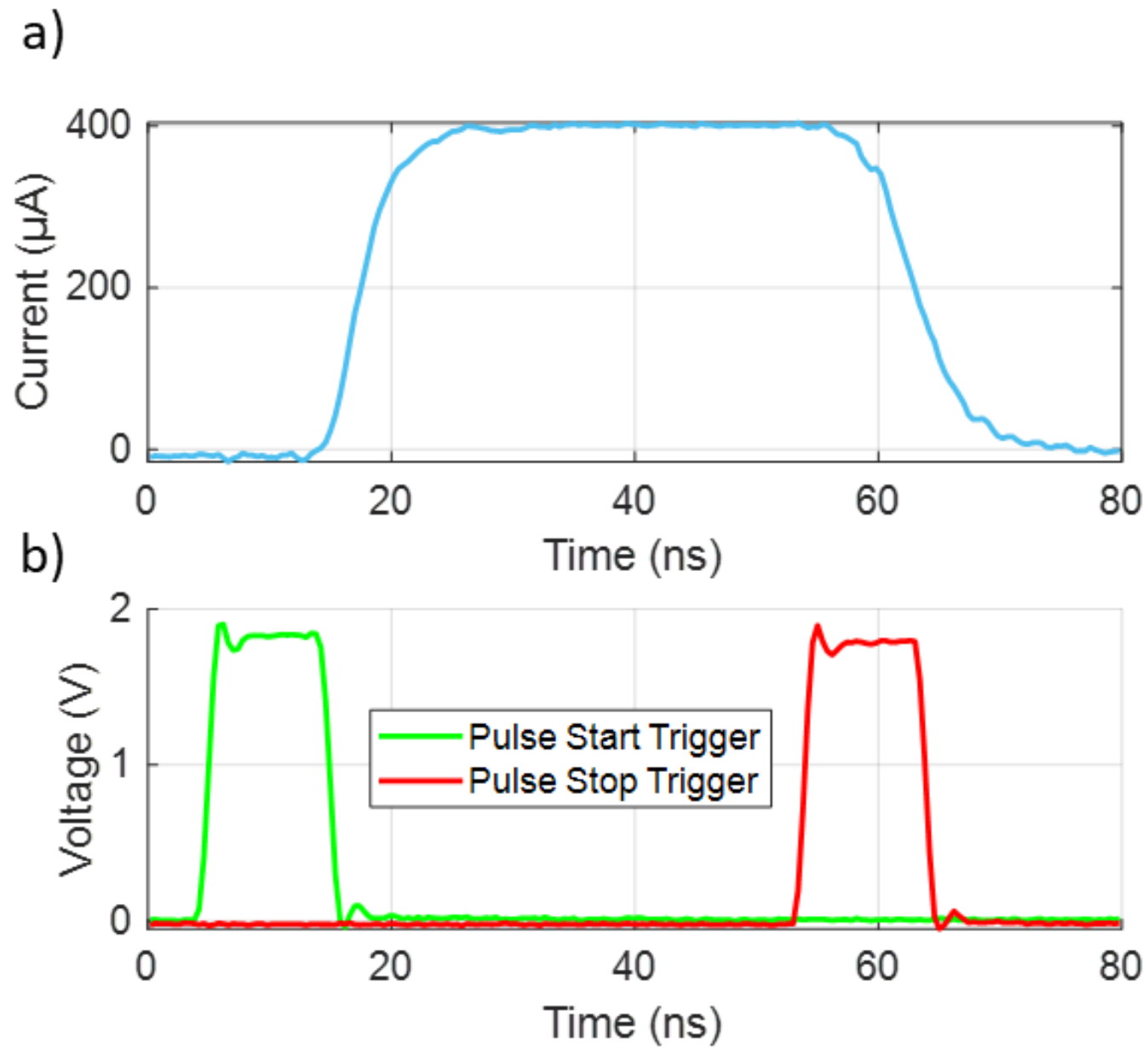

Figure 14 – a) Nominal current measurement of the analog output of the ASIC DUT at room temperature with the pulse width set by the start and stop triggers. b) The start and stop trigger pulses. Together, these graphs show the response time of the output waveform to the digital control signals.

The ASIC DUT's analog output pulse width is tunable, set by the falling edge of a pulse start trigger and the rising edge of a pulse stop trigger, as shown in Figure 14, allowing for a wide range of possible pulse widths with the resolution set by the capabilities of the AWG. The pulse height is also tunable, set by a 5-bit digital configuration data sequence that's processed by the digital block of the ASIC. All 31 states after state 0 were swept for the three mux states, and we found the pulse amplitude to have a range of 167 $\mu A \pm$ 11 $\mu A$ to 488 $\mu A \pm$ 14 $\mu A$, with an average step size of $10\ \mu A \pm 10\ \mu A$, as shown in Figure 14. All three Mux state outputs were tested and all were found to have approximately the same range and step size at room temperature. The rise and fall time of the pulse is also configurable, though this configurability was not tested at room temperature.

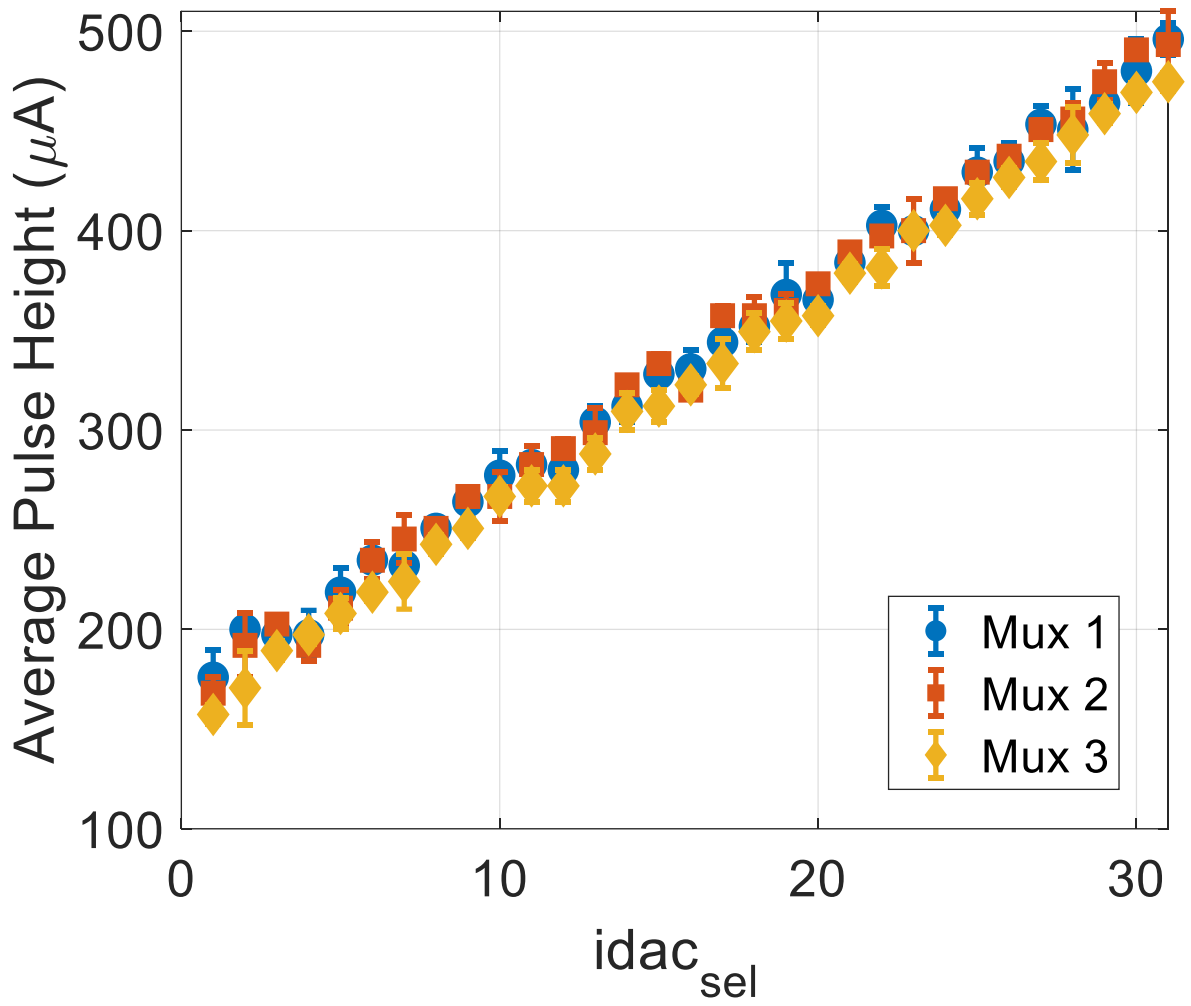

Figure 15 – Measurements of the average pulse height, with three pulses per data point, for three mux states at room temperature. The measurement indicates that the output amplitude is consistent across the different experiments (referred to as Mux 1, 2, and 3), in agreement with simulations shown in Figure 8a. Here $idac_{sel}$ is the digital control name used for the control of the pulse amplitude.

## B. Cryogenic Test Configuration and Measurements.

The ASIC was then mounted to a package within the milliKelvin stage of a Bluefors® XLD1000 as shown in Figure 16, below. The setup used identical test equipment to the room temperature setup as previously depicted in Figure 13. Different from the room temperature setup, the cryogenic setup includes approximately three meters of cable length between the test equipment and the chip; this cable length is a physical fridge wiring constraint for the measurement of the standalone ASIC DUT.

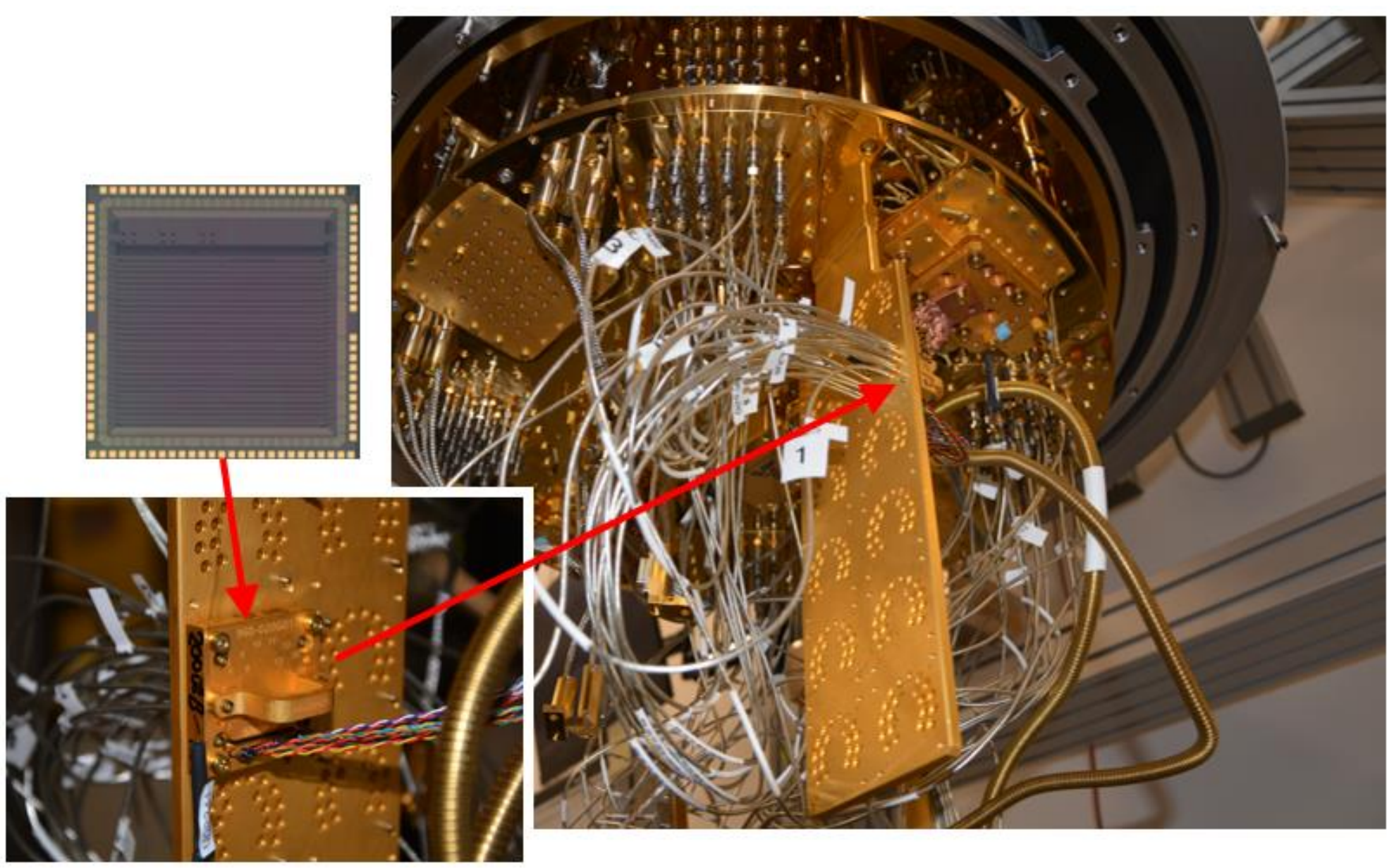

Figure 16 – The package for the ASIC DUT mounted within a Bluefors® XLD1000's milliKelvin stage.

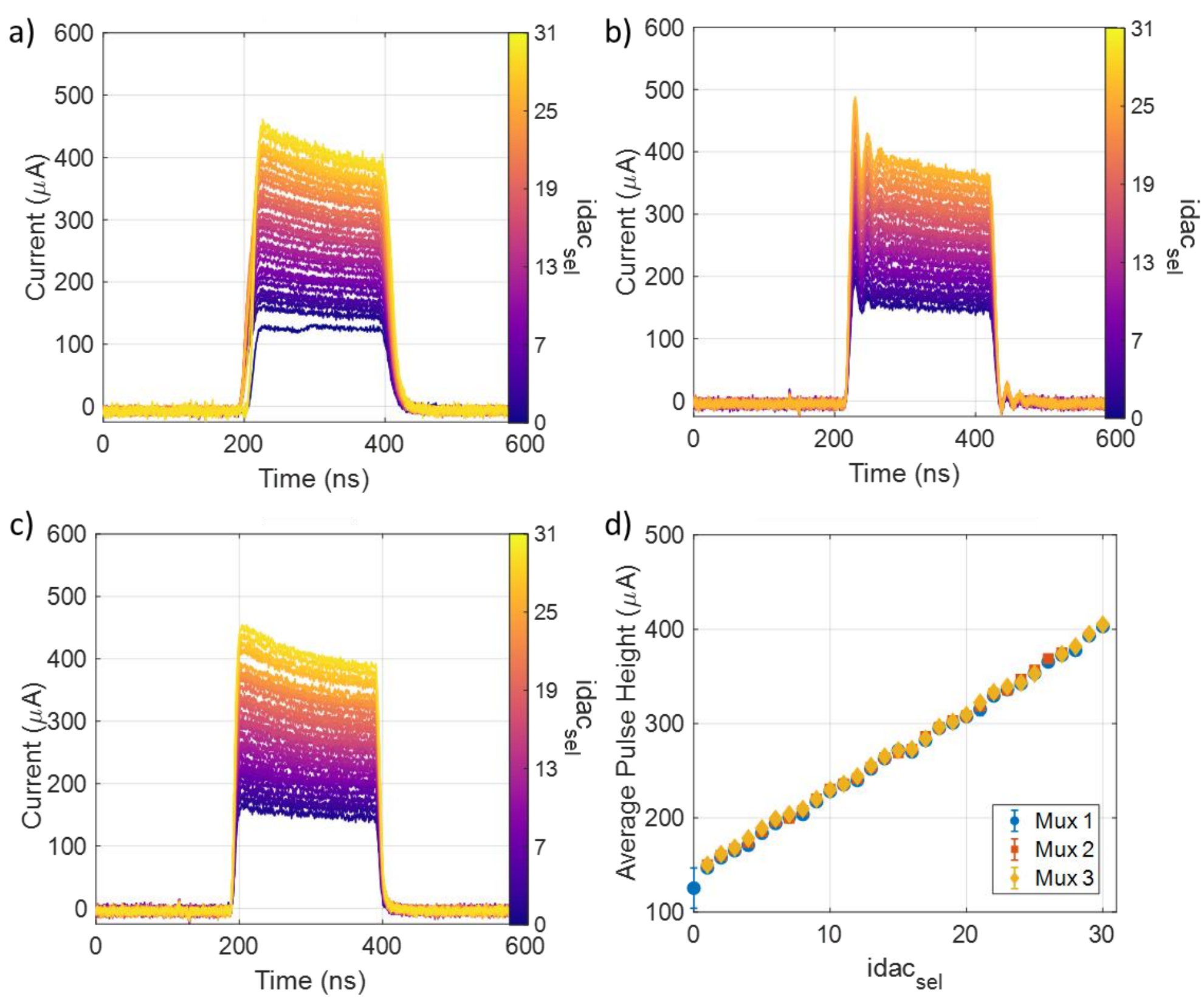

Figure 17 – Sweeping through current pulse amplitudes at 600 mK for each of the three identical on chip experiments, with a) the output pulses of mux state 1 for each idac$_{sel}$ state, b) the output pulses of mux state 2 for each idac$_{sel}$ state, and c) the output pulses of mux state 3 for each idac$_{sel}$ state, and d) the three average pulse height for the ouputs of all three mux states as a function of the state of idac$_{sel}$,

which ranges from 0 to 31. Again, $idac_{sel}$ is the digital control name used for the control of the pulse amplitude.

We tested the pulse height amplitude configurability at 600 mK, as shown in Figure 17. We found the output of all three mux states to approximately the same as show in Figure 17d. We found the average analog output pulse height to range from 130 $\mu A \pm 20\ \mu A$ to $403\ \mu A \pm 2\ \mu A$, with the majority of the error occurring in the 0 $idac_{sel}$ configuration state, with an average error of 2 $\mu A$ for the rest of the output states, and an average step size of 9 $\mu A \pm 4\ \mu A$, as shown in Figure 17d. We found the settling time for the pulse to increase as the amplitude increased, as shown in Figure 17a-c, with significant ringing in the pulse only in mux state 2, as shown in Figure 17b. The results fall within the expected range as shown by the simulations in Figure 8a.

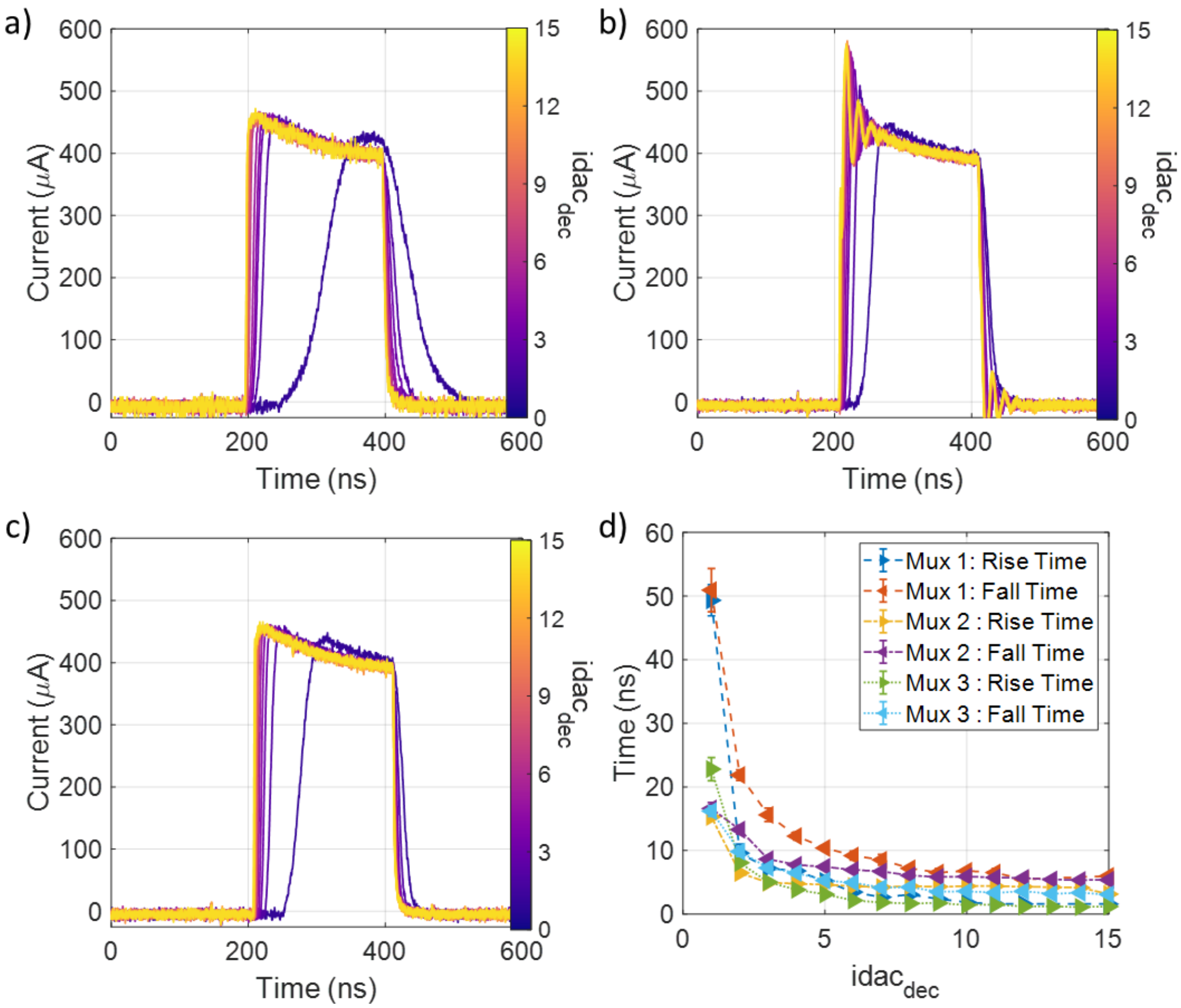

Figure 18 – Sweeping rise and fall time state configurations for the three identical experiments on chip, with a) the output pulses of mux state 1 for the various $idac_{dec}$ states, b) the output pulses of mux state 2 for the various $idac_{dec}$ states, and c) the output pulses of mux state 3 for the various $idac_{dec}$ states, and d) the rise and fall times as a function of the $idac_{dec}$ state as it ranges from 0 to 15, with rise times shown by the right facing triangles and fall times shown by the left facing triangles. Here $idac_{dec}$ is the digital control name used for the control of the rise and fall time.

The rise and fall configuration bits were stepped through with the dilution refrigerator held at 600 mK, as shown in Figure 18. The pulses were held at 400 μA with 200 ns pulse widths. The rise times range from 49 ns to 1.6 ns for Mux 1, 15.3 ns to 3.4 ns for Mux 2, and from 23 ns to 1.2 ns for Mux 3, as shown in Figure 18d. The fall times ranged from 51 ns to 6 ns for Mux 1, 17 ns to 5.4 ns for Mux 2, and 16 ns to 3 ns for Mux 3. These results fall within expectations as given by the simulation in Figure 8b.

## IV. CONCLUSIONS.

This paper presents an in situ current control solution for manipulating superconducting flux qubits or other similar superconducting circuits operated via analog flux control in a cryogenic environment. The demonstrated CMOS current DAC is controlled by a simplified digital interface using a minimum number of room temperature control lines that can be used to control multiple multiplexed devices. Implementation of the ASIC was achieved using tailored CMOS mixed signal design techniques on a conventional 90nm planar CMOS process using only early-stage cryogenic models of the available MOSFET devices and resistors.

While the first design iteration was measured to closely match circuit predictions, more work is needed to characterize the second iteration which has been sent for fabrication. In simulation, the second iteration achieves state-of-the-art gate-fidelities for cryo-CMOS controlled qubits. In future applications, one important design aspect would be related to thermal relief, packaging, and co-integration with quantum circuits and work towards implementing a solution at scale. Future work should focus on improving the state of cryogenic models across multiple semiconductor foundry offerings and increasing overall CMOS current DAC channel density per ASIC device with the aim of increasingly complex system demonstrations.

### *Authors:

A. Pelteku, D. Reitz, M. Covington, A. Frederick, H. Adebayo, K.D. Hillaire, N. Mitran, Z. Steffen, R. Schwartz, Z. Stegen, J.X. Przybysz, J. Yager, C. Walter, T. Graves, D. Meyer, J. Clark, G.R. Boyd, and A. Pesetski

## Acknowledgements

Spectre® is a registered trademark of Cadence Design Systems®, Inc. Tektronix® and Keithley® are registered trademarks of Tektronix®, Inc. Bluefors® is a registered trademark of Bluefors®, Inc. SkyWater® Technology is a registered trademark of SkyWater® Technology, Inc. GlobalFoundries® is a registered trademark of GlobalFoundries®, Inc.